# Distributional Consequences of Political Freedom: Inequality in Transition Countries

MONIKA WESOŁOWSKA
Department of Macroeconomics and Development Research, Poznań University of Economics and Business, Poznań, Poland

SŁAWOMIR KUŹMAR
Department of Macroeconomics and Development Research, Poznań University of Economics and Business, Poznań, Poland

BARTOSZ TOTLEBEN
Department of Macroeconomics and Development Research, Poznań University of Economics and Business, Poznań, Poland

DAWID PIĄTEK
Department of Macroeconomics and Development Research, Poznań University of Economics and Business, Poznań, Poland

**Abstract:**
The article addresses the origins of income inequality in post-socialist countries from Central and Eastern Europe and Central Asia, from 1991 to 2016. The aim is to analyze the relationship between democracy and income inequality. In previous studies, this topic has led to ambiguous findings, especially in the context of the group of countries we are focusing on. We examine whether the process of democratization co-occurred with changes in income distribution over the entire period under study, and its impact on individual income deciles to determine who benefited most from the new system.The obtained results allowed us to confirm that the actual relationship between democratization and income inequality did not exist, or at most was illusory in the 1990s, but it was present, relevant, and had a pro-equality character between 2001-2016. During that period, the development of the democratic system benefited at least 80% of the lower part of the income distribution, at the expense especially of the top decile's share of total income. Those results confirmed that democratization positively affected the shares of lower income deciles in post-socialist countries.

**Keywords:** income inequality, post-socialist countries, democratization, political freedom, distribution of income

# 1. Introduction

The influence of democracy on income inequality within society is of paramount importance, serving as a critical factor in shaping economic policy. Its potential to promote a more equitable distribution of income and to safeguard vulnerable populations is widely recognized (Acemoglu et al., 2015; Milanović, 1998; McCarty & Pontusson, 2009; Förster & Tóth, 2015). While the prevailing view supports the positive impact of democratization on inequality, literature also describes instances where an extremely different system, known as autocracy, might offer advantages for certain segments of society with lower incomes (Beitz, 1991).

Despite extensive research efforts, establishing a definitive relationship between income distribution and democratization, by which we consider the process aimed at transforming the non-democratic form of government into a democratic system, remains inconclusive. The volume of empirical work supporting the positive influence of democracy on income inequality is comparable to the amount of research pointing to the opposite thesis. Empirical confirmation of any relationship is extremely challenging, stemming not only from the unobserved heterogeneity within large study groups but also due to the limited availability of detailed income distribution data over extended periods. This ambiguity perpetuates the importance of studying this issue. Its inconclusiveness impacts both theoretical inference and empirical investigations into the correlation between democratization and inequality, resulting in significant research gaps.

The challenge becomes particularly apparent when post-socialist countries are included in the sample, exacerbating the existing gap. This was evidenced, among other things, by Bergh and Bjørnskov (2021). Our article addresses the question surrounding the origins of inequalities, analyzing the interplay between democracy and income inequalities in these post-socialist nations from Central and Eastern Europe as well as Central Asia. These countries, having transitioned to democracy around 30 years ago, exhibit a unique situation. At the beginning of the 1990s, they tried to replace the authoritarian systems with democracies, yielding varied results. Some of them have successfully established democratic and market-based systems, whereas others created authoritarian states with a limited scope of economic freedom (Piątek, 2016).

However, even when post-socialist countries transitioned to more democratic regimes, there are many examples where this change led to an increase rather than a decrease in inequality, contrary to expectations. We focus on political freedom as the institution that influences other political and economic institutions. In order to assess the relationship between political freedom and inequality in the studied countries, we conduct a series of analyses focused on the occurrence of correlations, direct and indirect effects of democratization on income shares of different deciles and inequality measures, basing our results on the existing literature.

Considering how difficult it is to study the relationship between democracy and inequality, due to the gradual nature of political freedom expansion and the simultaneous changes in other inequality determinants, like globalization, technology, and external shocks, we focus on the dynamic interplay and co-occurrence of these factors over time, rather than attributing causality to a single factor. This perspective allows us to come up with more certain conclusions and is sufficient to answer the questions we have been pursuing.

This paper contributes to the field by exploring the impact of democratization on income inequality, particularly within the context of post-socialist countries during their economic transition. This setting challenges the prevailing assumption that democracy diminishes income polarization. Our study also advances research on income distribution. The structure of this paper is as follows: Section 2 reviews existing literature on the consequences of democracy for inequality. Sections 3 and 4 present data on democracy and income inequality within transition countries, alongside the empirical strategy. Section 5 outlines our analytical approach, while Section 6 discusses the results. The last section concludes the article.

## 2. Democracy and Inequality: Literature Review

Literature distinguishes income inequality sources into endogenous and exogenous categories (Malinowski, 2016, p. 33). Certain income inequalities stem from intrinsic human qualities—endogenic factors—like intelligence, personality, charisma, and power. However, income inequalities may also arise from the social, political, and economic environment. In such cases, inequalities result from exogenous factors such as technological progress, economic processes (like globalization, financialization), and institutional frameworks (Roine & Waldenström, 2015). Our study focuses on a specific exogenous factor of institutional nature, namely the level of democratization, and its potential impact on inequality levels and income distribution.

Economic systems operate within a broader political framework, and how the political system influences income inequality hinges on the institutions and policies it establishes. These political constructs are shaped by the power distribution in society and how political institutions align preferences through mobilized interests (Acemoglu et al., 2005; Acemoglu & Robinson, 2014).

Acemoglu et al. (2005) and Acemoglu & Robinson (2014) emphasize the significant correlations between political and economic institutions, which play a significant role in determining economic outcomes and inequalities. They also present a model that demonstrates the relationships between political and economic institutions, economic results, and income distribution. Specifically, Acemoglu et al. (2005) highlight that economic institutions shape the incentives influencing economic entities, thus affecting the economic growth rate. However, these economic institutions are not exogenous, but endogenic. They are shaped by the entities that exercise political authority in a given country. Those who hold political power decide about economic institutions and, consequently, about economic results and income distribution.

However, political institutions—the regulations that govern incentives—ultimately determine who holds power (Acemoglu & Robinson, 2014). These institutions decide how authorities are elected and who may lawfully hold power. Political institutions are shaped by entities that rule the country. Besides de jure authority, there is also de facto authority from amassed wealth rather than legally binding mandates. Acemoglu and Robinson (2014) introduced a classification into extractive institutions, which hinder economic growth and worsen inequalities, and inclusive institutions, which increase citizens' wealth and reduce inequalities. Extractive institutions concentrate power in a narrow group, typical of nondemocratic regimes. Inclusive institutions distribute power broadly, ensure democratic elections, and uphold the rule of law.

According to existing literature, several theoretical mechanisms highlight how political institutions can impact income inequality. Democracy is expected to foster an equalizing effect. Citizen participation should prompt governmental actions that ensure the equitable sharing of both hardship and profits (McCarty & Pontusson, 2009). As highlighted by Lenski (1966), the spread of voting rights should lead to a more equal income distribution. Meltzer & Richard (1981) pointed out that poorer households tend to vote for politicians advocating higher redistribution and higher taxes, thus reducing inequalities. Also, in democratic societies, the median voter determines the tax rate based on income and poor groups can benefit more if tax revenues are distributed equally to all (Topuz, 2022). Additionally, reducing human capital variation through public education should also reduce inequality. Conversely, institutions that centralize political power within a narrow segment, as seen in non-democratic regimes, are expected to generate greater inequality (Acemoglu et al., 2015). Such regimes formulate policies benefiting the politically influential at the expense of other sectors of society. Economic institutions are used to generate economic rents which are appropriated by those in power. For example, regulations on occupation and residential choices disproportionately reduced wages for black Africans in Apartheid South Africa before 1994 (Wilse-Samson, 2013). What is likewise important, the literature also points to the possibility of a contingent relationship between democracy and inequality, claiming that the effect of democracy on inequality is contingent on the level of inequality at the time of democratization (Dorsch & Maarek, 2019; Bahamonde & Trasberg, 2021).

Despite these compelling arguments, literature has not reached a consensus on the relationship between democracy and inequality. Even within democratic political systems, substantial variation in economic inequality indicators exist, particularly among affluent nations. Furthermore, some advanced democracies have witnessed receding government efforts to alleviate inequality, coinciding with economic forces that encourage less equitable distribution of rewards (McCarty & Pontusson, 2009). Moreover, numerous instances demonstrate that the transition to a more democratic regime can result in increased, rather than decreased, inequality (Förster & Tóth, 2015). One of the most striking cases are Central and Eastern European countries during post-communist transitions, which encountered one of the most rapid and significant surges in inequality ever documented (Milanovic, 1998). Beitz (1982) even posits that authoritarian regimes might better protect the interests of the poorest social groups, albeit at a cost of holding power, and Paglayan (2020) adds that that democracy does not induce lower income inequality, more progressive taxation or pro-poor social policies.

As Sirowy & Inkeles (1990) pointed out, the number of studies supporting the hypothesis of positive influence of democracy on inequalities is balanced by the number of studies affirming the opposite. Gradstein and Milanovic (2004) have argued that the cross-national empirical evidence on democracy and inequality is ambiguous and not robust. Scheve and Stasavage (2009, 2010, 2012) indicated that democracy has limited influence on inequality and policy among OECD countries, and Gil et al. (2004) claimed that there is no discernible relationship between democracy and any policy outcome across different countries. Acemoglu et al. (2015) found a robust positive effect of democracy on tax revenues as a fraction of GDP, but no consistent impact on inequality. Bahamonde & Trasberg (2021) even find that democratic governance and high state capacity together cause higher levels of income inequality over time, through positive effects on foreign direct investment and financial development.

Wiseman (2017) demonstrated that enhanced economic freedom results in higher rates of income growth for the bottom 90% compared to the top 10%, leading to a more egalitarian income distribution. Trinugroho et al. (2023) find that democracy helps reduce inequality as it opens up possibilities to get more education for marginalized groups, implying higher income for those people. It is further indicated that with a lower level of democracy, the negative impact of inequality on economic growth may be higher (Islam, 2018), exacerbating social stratification.

Acemoglu & Robinson (2000) indicate, based on the United Kingdom, that democratization raises redistribution and spreads education; these factors decrease inequalities. Rodrik (1999) presented evidence linking democracy to a higher share of employee income in GDP. Li et al. (1998) showed an index of civil liberties is negatively correlated with inequality (greater civil liberties, lower inequality). According to Lundberg & Squire (2003), democracy increases the share of total income of the lowest quintile. Lindert (2004) provided evidence from OECD countries linking democratization to public spending, particularly on education; Persson and Tabellini (2003) presented similar cross-national evidence. Tavares & Wacziarg (2001) contend that democracy fosters growth by improving human capital accumulation and, less robustly, by lowering income inequality. Reuveny & Li (2003) studied both economic openness and democracy on income inequality. Their analysis of 69 countries from 1960 to1996 revealed that democracy and trade reduce income inequality. Roine & Waldenström (2015) demonstrate that in the long run, democracy (along with high marginal tax rates) is inversely correlated with the highest income shares. Balcázar (2016) analyzed the link between democracy and inequality in Latin America, uncovering that the degree of democracy experienced by birth cohorts during formative years relates to the dispersion of labor income in adulthood. Cohorts exposed to higher levels of democracy show lower income inequality.

Literature lacks consensus concerning the relationship between democracy and inequality. Theoretical indications suggest that democracy can enhance income equality directly or indirectly. However, counterarguments suggest that democratization may not necessarily benefit the poorer segments of society. In empirical terms, uncertainty remains regarding the existence of the indicated relationship and its direction, whether positive or negative, in income distribution equality. Challenges arise from cases deviating from the commonly observed patterns, like transition countries, which can distort conclusions when analyzing a large group of countries over an extended period.

### 3. Democracy and Inequality in Transition Countries

Analyzing the evolution of inequalities and their possible relation to pro-democratic reforms in transition countries is especially challenging. To begin with, it is difficult to precisely pinpoint how income distribution evolved in Western countries compared to countries in Central and Eastern Europe and Central Asia during the socialist era. This is due to variations in definitions, data quality and general differences between countries from both economic worlds, which make comparisons of the pre-1989 period very complicated (Flemming & Micklewright, 2000). Furthermore, these countries were diverse in terms of initial conditions, scale, and pace of the reforms being conducted. Nevertheless, the transformation period yielded many interesting observations and may help in at least partially understanding the puzzle surrounding the relationship between inequalities and democratization processes. In this section, we will explain the nature of the transformation to market economies, and examine data regarding both aspects that this article focuses on, with an awareness that any potential relationship between them, especially in the beginning of the transition, may stem from the inherently abrupt nature of transformation.

Before transformations, income inequalities were relatively low in socialist countries, primarily stemming from ideological motives that aimed to flatten the pay structure (Bukowski & Novokmet, 2017; Novokmet, 2017). Additional contributing factors included low registered unemployment, measures to prevent intergenerational transfer of private assets, price subsidies, rationing, and non-wage remuneration. Moreover, socialist states boasted a higher percentage of working women compared to Western countries (Flemming & Micklewright, 2000). Nevertheless, it is essential to note that despite low-income inequalities, pronounced consumption disparities existed due to scarce and regulated access to commodities. Obtaining rare products relied on personal connections, party membership, and hierarchies within the communist party (e.g., exclusive stores accessible only to party activists, security personnel, and high-ranking officials).

In the 1990s, income inequality underwent significant transformations; it not only increased but also became more diverse, a phenomenon unprecedented in other regions (see Figure 3.1) (Milanovic, 2001). Over the decade, the average Gini coefficient—a measure of income inequality—rose by 0.10 points, indicating a rapid shift towards greater inequality in these countries compared to their Western counterparts. This increase was particularly pronounced in former Soviet Union countries, while countries that later joined the European Union experienced milder changes in this area (Alvaredo & Gasparini, 2015). Specific outliers like

Georgia, Tajikistan, and Turkmenistan exhibited persistently high levels of inequality during and after the transition, although data quality requires cautious consideration. Additionally, during the initial transitional phase, income inequality disparities were narrower than at the end of the studied period, with the average top 10% income share being 0.24 in 1991, and 0.31 in 2016. At the same time, Gini coefficients were below 0.3 at the beginning of the transformation and averaged around 0.4 in 2016, with variations among countries.

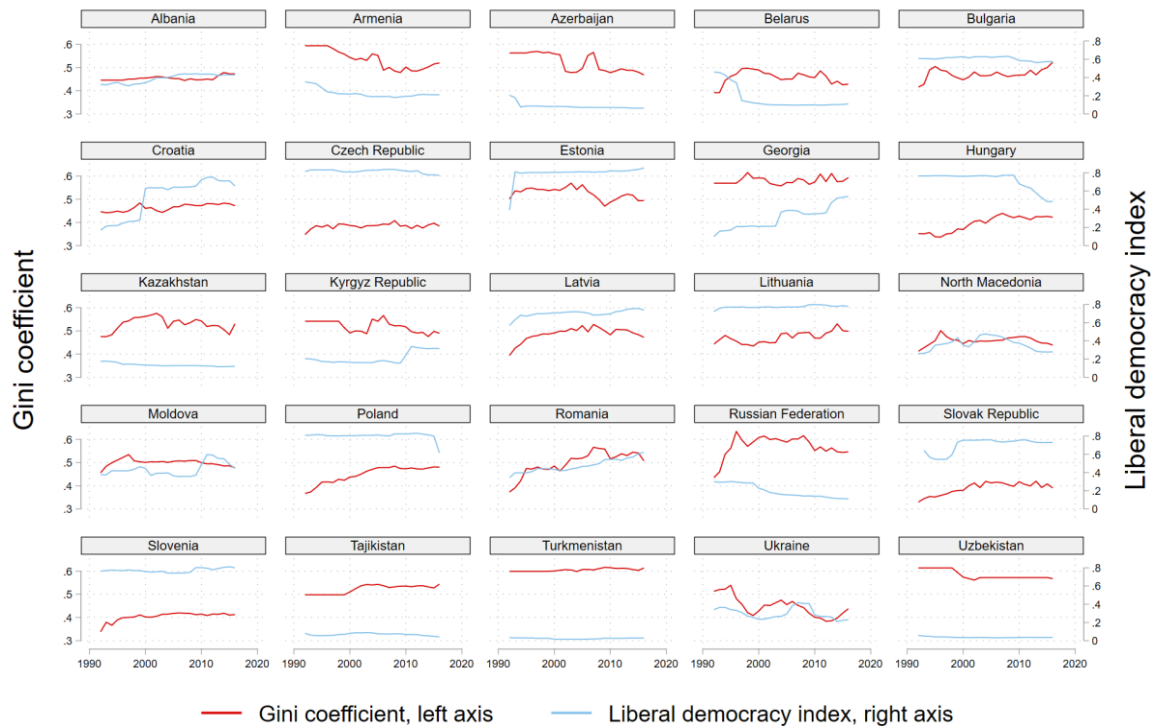

**Figure 3.1 Inequality levels and liberal democracy index in transition countries (years 1991-2016)**
Source: own elaboration based on: WID.WORLD database.

From a political standpoint, the transitions of the 1990s aimed to replace authoritarian systems with democracy and shift from centrally controlled economies to market economies (see Figures 3.2 and 3.3). Abandoning socialist ideology, which allowed the government to set salary levels, led to increased income. The introduction of market economy principles and privatization of state-owned companies further contributed to income inequality by increasing profit dispersion compared to compressed wage structures in state-owned firms (Alvaredo & Gasparini, 2015). Additionally, shifts in the political system led to the implementation or increase of utility fees for municipal services (Milanovic & Ersado, 2010) and a weakening of the minimum wage, impacting the poorer segments of society (Standing & Vaughan-Whitehead, 1995). The ongoing changes may therefore indicate a higher level of democratization, but the situation of the lowest-income groups worsened, contributing to an increase in the levels of income inequalities. This effect can be considered inevitable to some extent, since equality was a fundamental principle of socialism, and capitalism presupposes the presence of unequal distributions. Considering this, income inequality during the transition must have increased by the very nature of this process. At the same time, these countries experienced democratization - which was as we mentioned one of the goals of their transitions. However, the relation between income inequality and democratization is not so simple. The observed growth of inequality was also influenced by the other concerning phenomena at the intersection of politics and economy. Countries that lacked a properly functioning democracy experienced limited control over politicians, leading to corruption and "state capture" or "crony capitalism" (Havrylyshyn, 2006; Sonin, 2018). Such conditions hindered both political freedom, economic development and increased income inequality. Moreover, all post-socialist countries faced a so-called transformational recession (Kornai, 1994) at this time, another channel influencing income inequality growth, rising unemployment, and increased demand for state aid. The disruptions were significant and in such large quantities that despite the transition goal of increasing political freedom, many post-socialist countries still exhibited limited or no political freedom in 2020 compared to 1991 (see Figure 3.3) (Freedom House, 1992; Freedom House, 2021).

Comparing Figures 3.1 and 3.2, it becomes evident that the relationship between political freedom and income inequality changes. Countries with high rates of democracy at the beginning of the transition were able to maintain or increase them over the period under study. For example, Slovenia, the Czech Republic, Hungary, and Slovakia simultaneously had some of the lowest inequality rates in the sample. They are also in the subgroup of the countries categorized as fully free, which could better address societal needs through social transfers, thereby preventing further inequality growth (Novokmet et al., 2018). In contrast, countries with the lowest rates of democratization, such as Tajikistan, Turkmenistan, Uzbekistan, and Russia, simultaneously had higher income inequality. Additionally, Belarus, where democratization indexes fell immediately after entering the transition path, promptly recorded a jump in income polarization. Although this is not a strict rule, Poland, which maintained high institutionalized democracy but experienced a decline in liberal democracy only at the end of the study period, recorded an increasing trend in inequality but still cannot be counted among the countries with the highest polarization. Conversely, despite a decline in the liberal democracy index in Hungary in recent years, income inequality has not been significantly affected. At least part of this correlation can be explained by the reason pointed out by Hellman (1998). Using the example of post-socialist countries, the author points out that the groups that benefit most from the early distortions of transforming economies are interested in continuing to maintain a situation that generates high private profits at significant social costs. This means that countries whose political systems were more inclusive of those who were "transformation losers" were able to adopt and maintain more comprehensive reforms that generated gains in a larger portion of society than countries that insulated themselves from social pressures. Considering this,, countries that started and maintained higher democratization indexes as early as the early 1990s were more likely to have a more equal distribution of income during and after the economic transition.

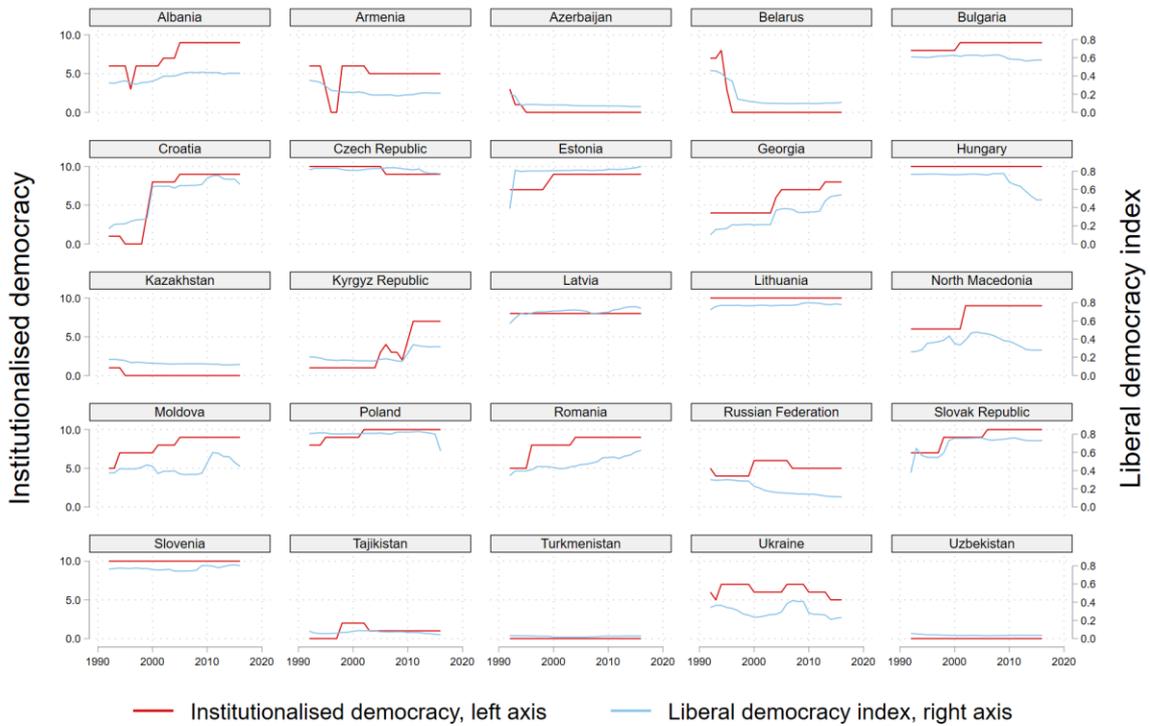

The institutionalized democracy scale (0 to 10) and the liberal democracy scale (0 to 1) are both positively oriented.

**Figure 3.2 Level of democratization in studied countries**
Source: own elaboration based on data presented in section 4.1.

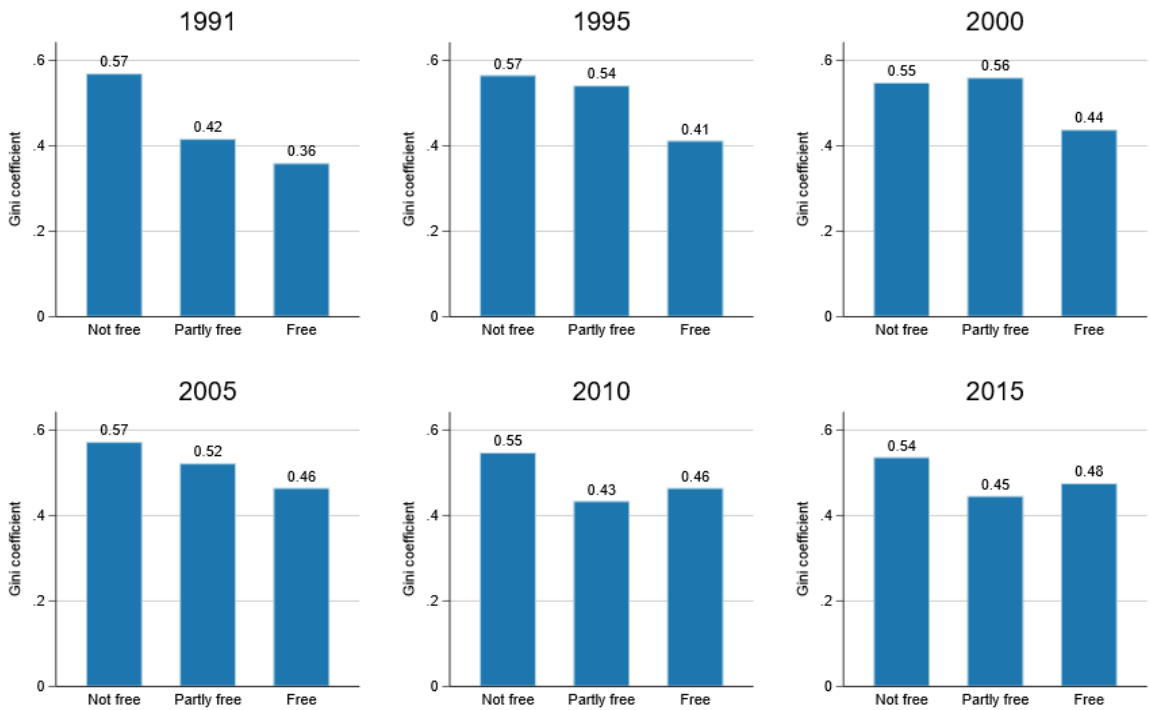

Mean values computed as population weighted average for each group of countries

**Figure 3.3 Average Gini coefficient in post-socialist countries by type of freedom**
Source: own elaboration based on data from Freedom House and WID.world database.

Summarizing the history of transformations and changes in income inequality, two dominant stages of the process emerge. The first decade, marked by intense and diverse changes that could significantly affect the entire research sample, may have yielded limited democratization effects, while the subsequent years represent a phase when democratization's influence on society might have been more discernible and direct. Initially, during the first decade of economic transition, post-socialist countries embarked on chaotic market reforms, incorporating various aspects of capitalism from different Western countries (Rapacki, 2019). In the early period, their aim was not to arrive at a chosen specific form of this system, but simply to build some system as close to capitalism as possible, but without choosing a specific type of it. In contrast, later on the economy and society were in a different place than in the early 1990s, and the changes that followed were more deliberate in terms of where exactly they would lead. For instance, some CEE countries began their path to European Union accession, which became a significant driving force for further development, particularly in aspects associated with structured democracy. Although after accession they did not remain immune to changes in income inequality levels (Brzeziński, 2018).

Based on the above, we decided that the relationship we were studying needed to be divided into two periods. As outlined earlier, the transition period in post-socialist countries remains problematic in empirical analysis. We assumed, based on literature, data, and preliminary analysis, that new institutions require time to develop before their effects manifest with a time lag. Moreover, throughout the transition interval, these countries were influenced by numerous other factors exerting direct impact on the surge in income inequality, among other outcomes. These factors stemmed from the shift toward market economies, inherently entailing heightened income polarization, concurrently with the cessation of artificial maintenance of flat income distribution. Consequently, this suggests that the initial rise of income inequality in post-socialist countries might not be inherently linked to the democratization process.

The adopted significant year for dividing the entire period is 2000, which was additionally confirmed from the initial tests conducted, showing that dividing the sample this way yields results that are significantly different between the sub-periods. Shifting this critical year, for instance to 2004, does not enhance the conclusions drawn.This is especially relevant given that the changes marking the initiation of post-socialist countries' integration into the EU began earlier, and 2004 symbolizes the confirmation of CEE member states as democratic market economies (Jarmołowicz, 2011). Roaf, Atoyan, Joshi and Krogulski (2014), on the other hand, distinguish between the 1990s, the early, and mid-2000s from a macroeconomic perspective. They highlight that before 2000, economic conditions were much more difficult and a high number of transition programs persisted in CEE countries. In contrast, the pace of reforms has slowed since the early 2000s. This means that the number of channels that can disturb our analysis results is lower in the 21st century. Given these considerations and the periods of significant change previously noted, we were further inclined to support the thesis that includes the year 2000 as part of the division of sub-periods.

The conducted literature studies, and the analysis of the empirical data illustrating inequalities and political freedom in post-socialist countries as well as a study of the diversity of the transition periods allowed for the formulation of the following research hypotheses.

H1: During the initial period of transition in post-socialist countries, there was no relationship between the democratization process and income inequality.

H2: After the initial phase, democratization had a positive impact on the shares of lower income deciles in post-socialist countries.

## 4. Data and Empirical Strategy

We aim to take a multidimensional approach to the relationship between income inequality and democratization in post-socialist countries. We do not only focus on general measures of inequality but also on income shares between deciles. Given the peculiarities of the countries studied and the period of economic transition described in an earlier chapter, we examine both the 1991-2016 period as a whole and its two main sub-phases. Despite time division, we avoid splitting countries. This approach stems from the fact that, despite the significant present-day differences among some countries, the problem under study remains the same for all of them. In analyzed articles, countries were not separated, allowing for generalized conclusions applicable to the entire group. Our study spans from 1991 to 2016, chosen due to the availability of the most comprehensive dataset encompassing explanatory variables in subsequent years. This period yields 650 country-year observations per variable across 25 countries, treated as separate time series data for each country and variable.

In the context of data, we used variables related to distribution of disposable income, along with information on the democratization of post-socialist countries, and additional control variables. Any missing values were handled through appropriate interpolation methods. First, we included data from the Lahoti et al. (2016) database - the Global Consumption and Income Project (GCIP). This dataset presents estimates in PPP units of monthly real income for every decile of the population. Using this information, we calculated the decile shares in total incomes.

We also integrate data from the GCIP database, using the Gini coefficient and Palma ratio as measures of inequality in our study. These measures served as dependent variables for our analysis. Including three measures of income inequality highlights diverse segments of the income distribution, enabling a comprehensive examination. The Gini coefficient (Farris, 2010) reflects overall population inequality; "1" indicates perfect inequality, while "0" indicates perfect equality, but identical values can reflect different income situations. The Palma ratio (Cobham, Schlögl, Sumner, 2016) compares the richest decile's income share to the poorest 40%, assuming stability in the middle-income range (Cobham, Sumner, 2013). Finally, decile shares detail the distribution's specific parts. This approach examines polarization changes, the relationship between extreme income groups, and individual subgroup dynamics, avoiding the limitation of focusing on one aspect and offering a fuller view of how democratization interacts with income inequality.

In our analysis, we included variables related to levels of democratization (and autocratization) in post-socialist countries. Democratization (autocratization) can be measured in various ways, taking into account their different aspects, which may not always be equally high or low. To comprehensively examine this phenomenon,, we decided to incorporate into the analysis several variables related to different areas. Moreover, we chose to use multiple sources of data to reduce the risk of bias stemming from single sources, which could influence the final outcomes.

We use datasets from databases such as Polity IV from the Center for Systemic Peace (CFSP), V-dem (version 13) and Worldwide Governance Indicators (WGI) as major sources of quantitative data on political systems. Due to high correlations between the data from the same source, we ultimately selected the five representative variables. From the WGI database, "Rule of Law" reflects perceptions of the extent to which agents trust and respect social rules, like the quality of enforcement of contracts, property rights, police and courts, and the likelihood of crime and violence in the country with the scale from -2.5 to 2.5. From the CFSP database, we derive the variable "Democratization Scale", which is on a scale of 0-10, indicating how democratized a country is, it considers a very wide range of criteria, like political freedom, media freedom, freedom of association, rule of law, fair elections, and more.

From the same source, we included the "Regulation of Participation", linked to governing the expression of political preferences in the context of existing political organizations. This variable employs a five-category scale, spanning from an unregulated state characterized by the absence of permanent national political organizations or systematic regime controls over political organizations through to a regulated state marked by relatively stable and enduring political groups vying for influence and positions, with minimal use of coercion and significant groups. The final source is V-dem, from which we chose the "Liberal democracy index" on 0-1 scale. This continuous index incorporates the liberal principle of democracy, emphasizing the safeguarding of individual and minority rights against both state tyranny and majority tyranny. Additionally, it also considers the level of electoral democracy. The final variable we incorporated is the "Equal Protection Index, which is a measure of the equality of social groups in respect for civil liberties or whether certain social groups are more advantaged. The index again adopts a 0-1 scale. "0" implies that members of some social groups have much fewer civil liberties than the general population, and "1" stands for the same level of civil liberties among all social groups.

Selected variables aim for broad coverage of democratization aspects. We introduce a very broad variable that is a general measure of the scale of democratization, and several smaller ones related to the role of law in the state and the extent to which it is respected and enforced, the method of shaping the political environment, the possibility of expressing political preferences and liberality connected with lack of tyranny and equal protection of all social groups. Each of the above variables consists of several aspects examined, and when selecting them we tried to maintain diversity, even though they all examine nearby areas. The incorporation of all of them allows for a comprehensive understanding of the impact of democratization on income distribution.

The final segment of our dataset encompasses control variables, with data sources from the World Development Indicators, Human Development Index, KOF Globalization Index, and International Monetary Fund. As we indicated earlier, democratization is not the only aspect that influenced changes in income redistribution, especially in the initial period of economic transformation that we presented in section 3. When post-socialist countries were just creating democratic standards (or consolidated autocratization), a lot of other changes affected many aspects simultaneously, which could also affect income in the society. We posit that, alongside democratization, income distribution can be influenced by improvements in the quality of life stemming from systemic shifts. In our study, we assume a general extension in the expected duration of education and life across the studied period. Moreover, political transformations, particularly those embracing a more democratic path, have co-evolved with the increasing globalization of countries and the heightened openness of their economies. This has presented additional opportunities, often higher wages, especially for the wealthier part of society. Consequently, we include variables such as the globalization index and data on exports and imports as a percentage of GDP. The final set of variables pertains to economic performance, represented through real GDP growth and the unemployment rate. The 1990s witnessed sporadic spikes in unemployment across all post-socialist economies, detrimentally impacting the lowest income brackets. This scenario could have potentially diminished the positive impact of democratization on this group. This is quite different from the case of GDP, which consistently exhibited an upward trend in all countries. In instances of data gaps, we applied interpolation by using the average dynamics of change or the mean value between the preceding and subsequent observations, depending on the nature of the variable.

Table 4.1 shows the basic statistics for the variables, for their better understanding. Due to the fact that the variables come from similar areas than before the analysis began, some of them were transformed accordingly. The goal was to reduce differences in the spread of values of a given variable to avoid disturbing the models, and to lower internal correlations to counteract the possibility of multicollinearity. Variables with larger data spans such as those from the globalization index were logarithmized, and in the case of variables related to quality of life and values from the democratization scale were square-rooted. Variables with smaller spreads, like import, were squared up. In addition, from the fact that regulation of participation is a categorical variable equal to one of the five values, it was split into four binary variables (the introduction of a fifth binary variable would distort the model, from the fact that its value would follow directly from the implemented variables). Figure 4.1 illustrates correlations among the 11 independent variables used. Some variables exhibit moderately high correlations, like globalization, expected years of education, life expectancy, and trade. However, this does not significantly affect the results. The correlations between these variables are not perfect, only two are above 70%, which ensures their independence within the study. These variables examine different aspects, enhancing result robustness. In addition, as indicated in the previous section, the study primarily focuses on examining the co-occurrence and correlation between the levels of inequality and democratization, rather than the direct effect of one aspect on the other.

|  | mean | std | min | 0.25 | 0.5 | 0.75 | max |
|---|---|---|---|---|---|---|---|
| income decile 1 | 0.025 | 0.009 | 0.001 | 0.019 | 0.023 | 0.030 | 0.058 |
| income decile 2 | 0.039 | 0.010 | 0.005 | 0.031 | 0.037 | 0.045 | 0.068 |
| income decile 3 | 0.050 | 0.011 | 0.013 | 0.041 | 0.050 | 0.058 | 0.076 |
| income decile 4 | 0.061 | 0.011 | 0.023 | 0.051 | 0.063 | 0.069 | 0.083 |
| income decile 5 | 0.072 | 0.010 | 0.038 | 0.062 | 0.074 | 0.081 | 0.090 |
| income decile 6 | 0.084 | 0.009 | 0.055 | 0.075 | 0.087 | 0.092 | 0.103 |
| income decile 7 | 0.099 | 0.008 | 0.074 | 0.091 | 0.101 | 0.105 | 0.123 |
| income decile 8 | 0.119 | 0.006 | 0.103 | 0.114 | 0.118 | 0.121 | 0.145 |
| income decile 9 | 0.152 | 0.008 | 0.130 | 0.148 | 0.152 | 0.156 | 0.196 |
| income decile 10 | 0.299 | 0.063 | 0.176 | 0.245 | 0.282 | 0.361 | 0.501 |
| Palma ratio | 1.886 | 0.885 | 0.620 | 1.236 | 1.593 | 2.570 | 10.699 |
| Gini coefficient | 0.376 | 0.078 | 0.182 | 0.323 | 0.372 | 0.448 | 0.577 |
| democratization scale | 2.169 | 1.15 | 0 | 1.499 | 2.645 | 3.0.0 | 3.162 |
| rule of law | -0.245 | 0.784 | -1.839 | -0.879 | -0.335 | 0.490 | 1.37 |
| liberal democracy | 0.432 | 0.397 | 0.016 | 0.17 | 0.397 | 0.741 | 0.851 |
| equal protection | 1.314 | 0.724 | -1.32 | 1.01 | 1366 | 1.678 | 2.583 |
| regulation of participation | 3.050 | 1.161 | 1 | dummy | dummy | dummy | 5 |
| GDP growth | 2.458 | 7.465 | -45.36 | 0.436 | 3.834 | 7.465 | 33.030 |
| years of life | 70.92 | 4.256 | 52.87 | 68.314 | 71.011 | 74.051 | 81.158 |
| years of education | 13.05 | 2.011 | 7.37 | 11.523 | 12.895 | 14.754 | 17.709 |
| export | 0.439 | 0.179 | -0.013 | 0.301 | 0.414 | 0.556 | 0.938 |
| import | 0.297 | 0.191 | 0 | 0.154 | 0.243 | 0.420 | 11.191 |
| globalisation | 4.048 | 0.2876 | 3.137 | 3.883 | 4.121 | 4.272 | 4.4457 |
| unemployment | 10.581 | 6.244 | 0.600 | 6.623 | 9.155 | 13.240 | 38.800 |

**Table 4.1 Basic statistics of variables**
Source: own elaboration

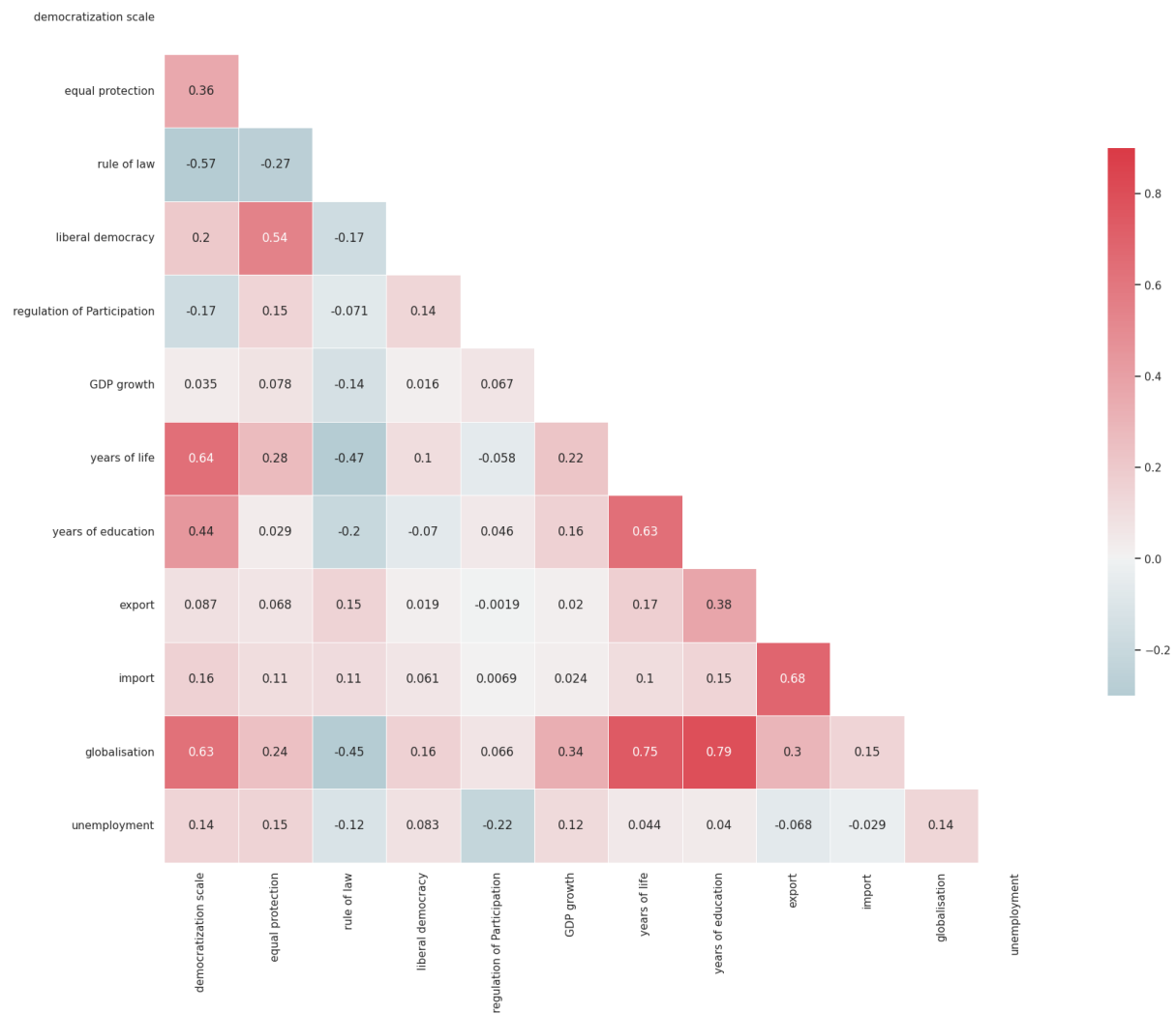

**Figure 4.1 Correlation between democracy and control variables**
Source: own elaboration based on data presented in section 4.1.

## 5. Analysis

To fulfill our scientific objectives, we conducted a series of panel regressions, in which the explained variables are previously introduced measures of inequality. We postulate that democratization interacts differently depending on its place in the income distribution, so we created panel regressions, for each decile, Gini coefficient, and Palma ratio. Regressions were run in several versions, with all democratic and control variables simultaneously, and a single democratic variable with all control variables. In addition, everything was recalculated for three time intervals (1991–2016, 1991–2000, 2001–2016). We chose the panel approach because it allows controlling for unobserved country-level heterogeneity that might affect the outcomes. We also include random effects (1) (Cameron & Trivedi, 1998), capturing country-specific intercepts, which account for time-invariant and country-specific characteristics. Random effects models were validated using the Hausman test, which assesses the appropriateness of those effects as estimators in comparison to fixed effects. Applying significance at the 5%, we accept the null hypothesis in favor of the random effects for all of the deciles, Palma ratio and Gini coefficient analyses. For more precise results we add robust standard errors to avoid the risk of obtaining incorrect standard errors if heteroskedasticity were present but not detected and we checked the correlations of the residuals with the regressors, whose low scores (all below 0.3) indicate that there is no risk of endogeneity in the model.

In the last part of the analysis, we supplement the results from the panel regression with complementary results for average marginal effects (AME) (2), calculated using earlier panel models with the same data. Introducing AME allows us to better understand the influence of individual variables on the dependent variable in the model. While panel regression shows the overall relationship between the independent and dependent variables, AME enables assessment of the specific effect of a change in a particular variable on the dependent variable, assuming other variables in the model remain constant, making them useful to represent causal effects in econometric applications (Aguirregabiria & Carro, 2021).

$$(1) \quad Y_{it} = \beta_0 + \sum_{k=1}^{K} \beta_k X_{k,it} + u_i + \lambda_t + \varepsilon_{it}$$

$$(2) \quad ME_k = \frac{\partial y_{it}}{\partial y_k} = \beta_k \cdot \underline{X_k}$$

where:
$Y_{it}$ - the dependent variable for observation *i* in time period *t*,
$\beta_0$ - intercept,
$\sum_{k=1}^{K} \beta_k X_{k,it}$ - the sum of the products of the coefficients $\beta_k$ and the explanatory variables $X_{k,it}$,
k=1,2,..,K - represents the index of the explanatory variables in the model,
$\beta_k$ - the regression coefficient for variable $X_k$,
$X_{k,it}$ - the value of the *k*-th explanatory variable for unit *i* at time *t*,
$u_i$ - the individual-specific effect (Entity Effect),
$\lambda_t$ - the time-specific effect (Time Effect)
$\varepsilon_{it}$ -Składnik błędu losowego
$\underline{X_k}$ - the random error term.

## 6. Results and discussion

Table 6.1 shows the panel regression results for two measures of income inequality - the Gini coefficients and Palma ratio. The analysis of those indicators was carried out for the entire period of economic changes from 1991 to 2016 and for the two sub-periods divided at the year 2000, which followed the period of initial economic transformation. The fit of the created models is moderately good in the case of the Palma ratio with R2 between 0.55 to 0.62, and very good for the Gini coefficient with R2 between 0.78 and 0.86. Importantly, in both cases, a better fit is found in the case of the second examined period (2001-2016).

For the entire studied period, at the 5% significance level, there was a negative direction of democratization`s impact on the level of income inequality indicated by the analyzes for both measures. The overall scale of democratization was significant for both variables throughout the whole period, but when divided into subperiods it turned out to be important only for the Palma ratio in the years 1991-2000. However, all tested components of democratization showed a significant relationship with the level of income polarization in the later period. At a more detailed extent, the significance and negative direction of the coefficient for the rule of law particularly occurring in after 2000, with very high significance for both measures. Considering that this variable often takes on negative values in the sample (positive values more often after 2000 than before) its low values are associated with increasing inequality and high values with decreasing it, all the more indicating its relevance to income equality. The results for equal protection referring to the equal distribution of civil liberties without threatened rights and freedoms of one social group by the actions of another one in the analysis for the entire period showed the negative impact on inequality. In the context of subperiods, this variable is significant only after 2000, but the direction of its coefficient is positive. Furthermore, the analysis on regulation of participation shows that throughout the period and the first sub-period, moving away from a fully unregulated state and creating more permanent political organizations is significant for the level of income inequality. The last variable related to democratization is the liberal democracy index, which pertains to the creation of inclusive institutions, and turned out to be significant only in the second analyzed period, with a pro-equality character.

The analysis of control variables allowed for the inclusion of additional dimensions of influence on income inequality. Firstly, over the entire period, the extension of the years of education in post-socialist countries had an egalitarian character, indicating the emergence of an educational rent (Woessmann, 2015) in the studied area. The trade situation (export and import variables) was characterized by balancing influences. Strengthening exports, related to greater production or entrepreneurship, constituting positive stimulation for the labor market and incomes, reduced the Palma ratio, but this effect reversed with higher imports. With equal growth in both categories, the overall impact remained negative for inequality. Globalization contributed to increasing income inequality, especially in the first period considered. After a gradual stabilization in the number of years devoted to education in the population, which had an inequality-reducing impact in 1991-2000, it ceased to be a significant variable in the 21st century. Moreover, the relationship between unemployment and inequality measures is surprising due to the negative direction of the coefficients. Considering that higher unemployment should lead to higher inequality due to the growing segment of society with low or no income, this relationship may result from the coexistence of the lowest levels of inequality at the beginning of the transformation. After the cessation of artificially maintaining full employment, unemployment rates increased dramatically, leading to the co-occurrence of extreme values of two factors, as explained in section 3.

For an in-depth analysis, we examined the average marginal effects[1] to assess how changes in the dependent variables would be shaped with respect to changes in individual democratic variables while keeping the other dependent variables constant. Figure 6.1 presents the obtained results, which deepen the conclusions already presented. The analysis showed that in the first period, only the scale of democratization and equal protection of social groups had reducing effects on income inequality. However, in the second period, for both measures, only the rule of law exhibited positive effects for inequality, although the magnitude of the effect decreased over the years.

Precisely, a one-unit increase in the democratization scale (or more accurately, a one-unit increase in the square root of the democratization scale value, since this variable has been transformed by square rooting, so its values vary between 0 and 3.16) in the second period could lower the values taken by the Palma ratio by 0.156, which, with the maximum value of the index after 2000 equal to 3.18, means a decrease of at least 4.1%, or a 0.5 percentage point decrease in the Gini coefficient. A unit increase in liberal democracy could lower the Palma ratio by 0.06 units and the Gini coefficient by 1.2 percentage points For equal protection, the impact was 0.02 or 0.8 percentage points for equivalent measures. In contrast, a one-unit increase in the variable of regulation of participation was associated with effects of 0.025 and 1.1 percentage points which, in the case of a situation where an increase from the lowest to the highest value taken by the variable occurred, would be associated with effects of 0.08 and 4.4

---

[1] The statistical significance of AME was tested based on the t statistic. The only statistically insignificant result was for the rule of law for the Palma ratio in the first subperiod. In all other cases, AME for Palma ratio, Gini coefficient, and income shares were significant at the 1% level.

percentage points In contrast, the effects of only non pro-equality variables were 0.08 and 0.9 percentage point

Our results are consistent with those of other authors. Findings for rule of law are in line with empirical analyzes such as Bhagat (2020) indicating a more equitable distribution of income at the occurrence of higher levels of this variable and theoretical ones regarding the importance of inclusive political institutions, like Acemoglu & Robinson (2014), of which the rule of law is an important component. Outcomes for civil liberties for the entire period are confirmed by the research of Li et al. (1998), which shows a negative correlation of civil liberties with inequality. Additionally, the fact that the democratization scale takes into account aspects related to electoral democracy confirms the conclusions of McCarty & Pontusson (2000). The authors pointed to a more equitable distribution of benefits in society with citizen participation that could encourage the government to act pro-socially.

In the case of control variables, the pro-equality impact of education, which is affected by the development of democratic governments, is due not only to the emergence of educational annuities, but also to the fact that, as Dahlum and Knutsen (2017) point out, democratization is positively related to aspects such as the percentage of people attending school and the number of years people remain in the education system. Moreover, the outcome for net trade is partly in line with Reuveny and Li's (2003) conclusions regarding the positive impact of foreign trade on equality.

On the other hand, according to Heimberger (2020), globalization progressing over time has contributed to increasing income inequality, which became apparent in the analysis. An explanation for such a relationship was provided by Jaumotte, Lall and Papageorgiou (2013). Globalization tends to relatively increase the demand for skills and education, and although it may bring benefits to the entire population by reducing poverty, these benefits are disproportionately higher for people who currently have higher education and skills. However, with more equal access to education, the benefits of globalization may also be more evenly distributed, which explains the lower impact of globalization in the second period considered after a gradual stabilization in the number of years devoted to education in population, which despite its inequality-reducing impact in 1991-2000, has ceased to be a significant variable in the 21st century.

This part of the analysis revealed that changes in democratization were significant in the model for income inequality throughout the entire period and were associated with lower levels of the Gini coefficient and the Palma ratio. However, in the later period of transformation (2001-2016), more aspects of the studied process were significant for the change in income than in the first sub-period (1991-2000), and additionally, the magnitudes of the coefficients were also higher in that period. Furthermore, the analysis of average marginal effects indicated that the examined variables, apart from the rule of law, had egalitarian significance, with the remaining variables holding constant more often in the 21st-century period. This suggests that democratization was a positive process for income equality after the time of greatest upheaval, when the new systems were being formed and democratization was evolving, and its effects were more evident after its initial shaping.

**Table 6.1 Cumulative regressions for Gini coefficient and Palma ratio**

|  | 1991-2016 | | 1991-2000 | | 2001-2016 | |
| --- | --- | --- | --- | --- | --- | --- |
|  | Palma | Gini | Palma | Gini | Palma | Gini |
| democratization scale | -0.302*** | -0.021*** | -0.326* | -0.009 | -0.055 | -0.002 |
| rule of law | -0.183* | -0.024*** | -0.395 | -0.033* | -0.422*** | -0.047*** |
| liberal democracy | 0.02 | -0.024 | -0.35 | -0.032 | -0.288* | -0.046*** |
| equal protection | -0.165*** | -0.007 | -0.171 | -0.006 | 0.007 | 0.008* |
| regulation of participation[=2] | -0.469* | -0.016 | -0.226 | -0.037 | 0.457** | 0.057** |
| regulation of participation[=3] | -0.41 | -0.017 | -0.857* | -0.071** | 0.584*** | 0.07*** |
| regulation of participation[=4] | -0.708** | -0.051** | -1.18** | -0.074** | 0.303* | 0.033* |
| regulation of participation[=5] | -0.408 | -0.013 | -0.063 | -0.022 | 0.486* | 0.053** |
| GDP growth | 0.004 | -0.00007 | 0.001 | -0.0005 | 0.001 | 0.0002 |
| years of life | -0.104 | -0.006 | -0.267 | -0.013 | -0.583*** | -0.014 |
| years of education | -0.853*** | -0.057*** | -1.151* | -0.089** | -0.356 | -0.018 |
| export | -0.686** | -0.022 | -1.063* | -0.047 | 0.176 | 0.016 |
| import | 0.536* | 0.025 | 0.601* | 0.04 | 0.122 | -0.013 |
| globalization | 1.817*** | 0.179*** | 2.494*** | 0.226*** | 1.872*** | 0.124*** |
| unemployment | -0.01** | -0.001* | -0.01 | -0.001 | -0.005 | -0.0003 |
| County RE | ✓ | ✓ | ✓ | ✓ | ✓ | ✓ |
| Period RE | ✓ | ✓ | ✓ | ✓ | ✓ | ✓ |
| Cov. Estimator | Robust | Robust | Robust | Robust | Robust | Robust |
| $R^2$ | 0.55 | 0.78 | 0.50 | 0.78 | 0.62 | 0.86 |

Signif. codes: 0.001 '***' 0.01 '**' 0.05 '*'
Source: own elaboration based on data presented in section 4.1.
**Figure 6.1 Average marginal effects for democratization variables from the models for Palma ratio and Gini coefficient**

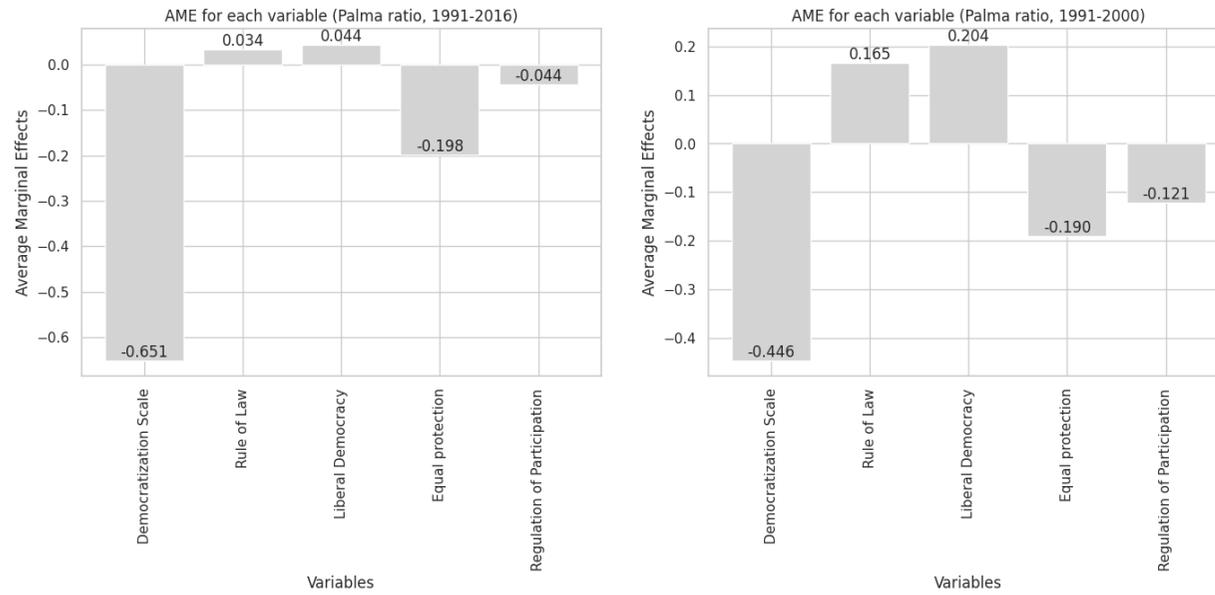

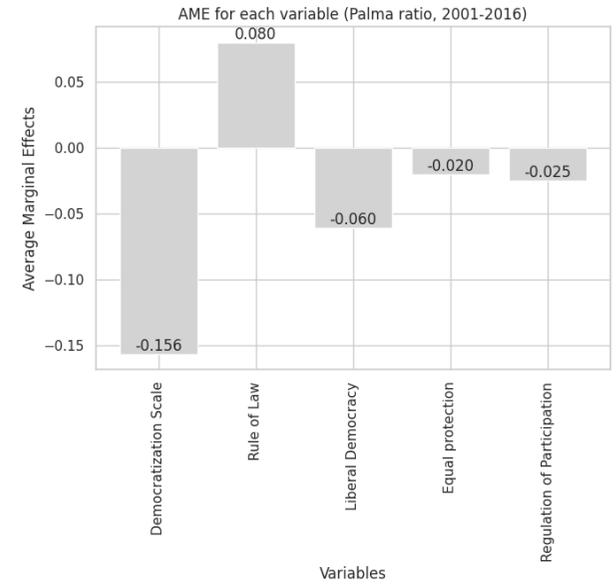
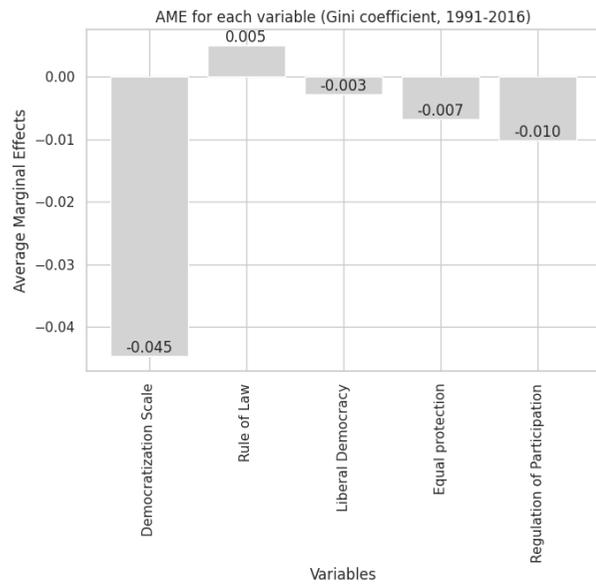
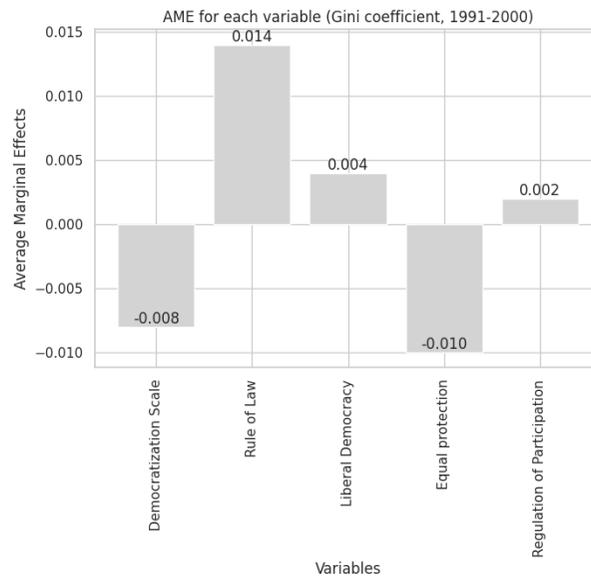
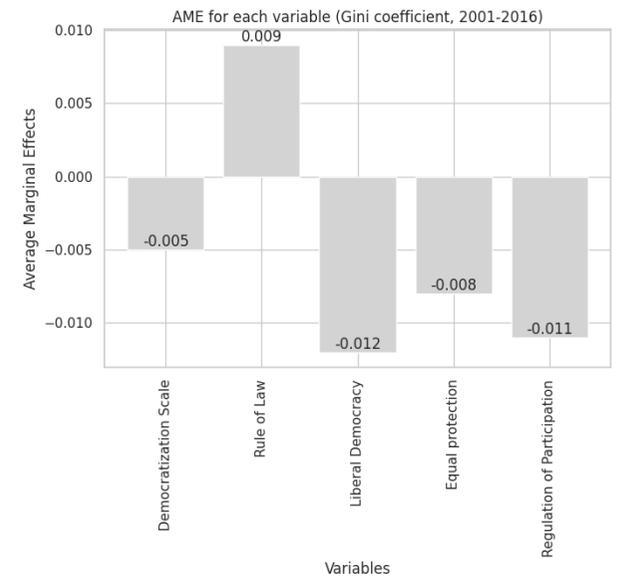

Before proceeding to further analysis, to better understand the relationship between income deciles, we analyzed their correlations. In figure 6.2, the significant variation in the interaction is noticeable between pairs of deciles. The shares of the poorer population show a strong positive relationship with each other but decrease with the highest shares held by the top 10%. We deduce that similar variation may result with independent variables affecting each decile. When comparing the bottom 80% and top 20%, the difference becomes clear, as the relationship is represented by lines with a 45-degree slope.

A more comprehensive analysis of the above relation reveals an interesting regularity. Figure 6.3 shows a very strong positive correlation between the initial eight deciles, contrasting with a strong negative correlation within the ninth and tenth deciles and remaining shares. More specifically, correlations between the first seven of all ten deciles are positive and strong or very strong, in most cases exceeding 90%, reaching up to 99%. This is especially true for correlations between deciles from the second to the seventh, where the lowest correlation is 70%. The lowest decile is very strongly correlated with the second and third deciles (95% and 90%), but this strength weakens with each successive decile. The correlation with deciles 5th and 6th is already strong (78% and 69%) and moderate with decile 7 (52%). Also breaking out of the general relationship between the top 80% of the income distribution is decile 8, for which there is only weak or very weak correlation with the two lowest deciles (4% and 22%). The same is true for the next few deciles, where the correlation is weak or moderately strong, though further positive, meaning that when the shares of decile 8 increase, they are correlated with not much positive change in the lower deciles. However, the correlation with the seventh decile is very strong at 84%, indicating the co-occurrence of large and unidirectional changes in both deciles simultaneously. The 9th decile has a very strong negative correlation with the first seven deciles, with a correlation of -82% with the 1st decile, -64% with the 4th decile, and only -18% with the 7th decile. This can be interpreted as an increase in the share of the 9th decile being associated with a decrease in income, especially in the lowest deciles. The only exception here is the 8th decile, with which the correlation reaches 37% indicating a very weak but positive relationship. On the other hand, an increase in the share of the richest decile comes at the expense of all deciles except the 9th (correlation equal to 49%), with the weakest correlation in the case of the 8th decile -61% and surprisingly the 1st decile at -78%, with the rest of the results ranging from -91% to -93%.

These observations indicate that increased income shares in the ninth and tenth deciles correspond to decreased levels of income shares in the other income groups, leading to higher levels of inequality, with a weak positive impact of the eighth decile, where growth does not lead to deterioration of the poorer part of the population. In addition, although improvements in the income distribution position of deciles two through seven and even eight are very strong, their relationship with decile one is weaker. This means that, in general, greater equality and wider access to income in those deciles translate less (or not at all) into positive changes in the income situation of the poorest.

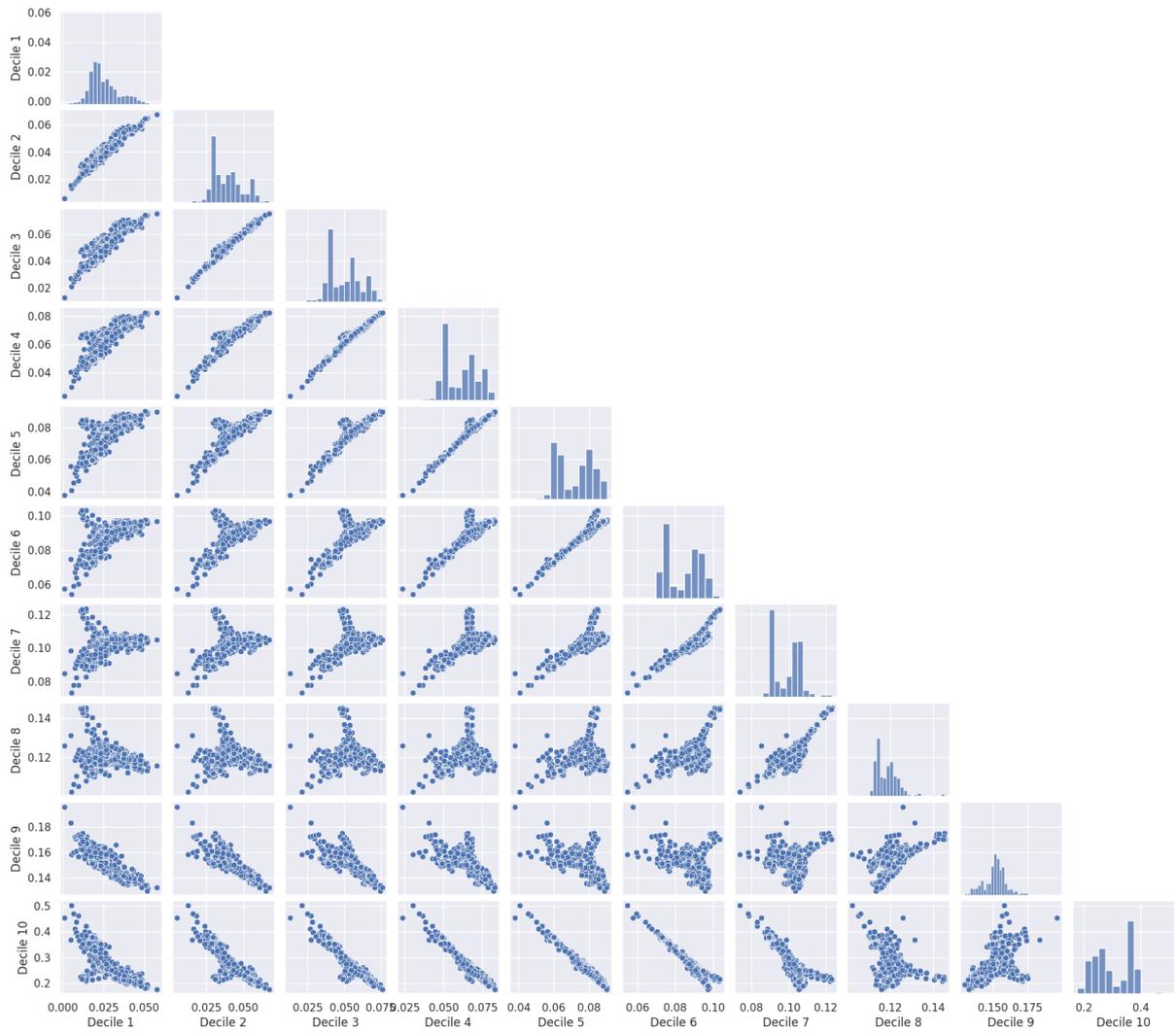

**Figure 6.2 Relations between income shares deciles - pairplot analysis**
Source: own elaboration based on data from GCIP database.

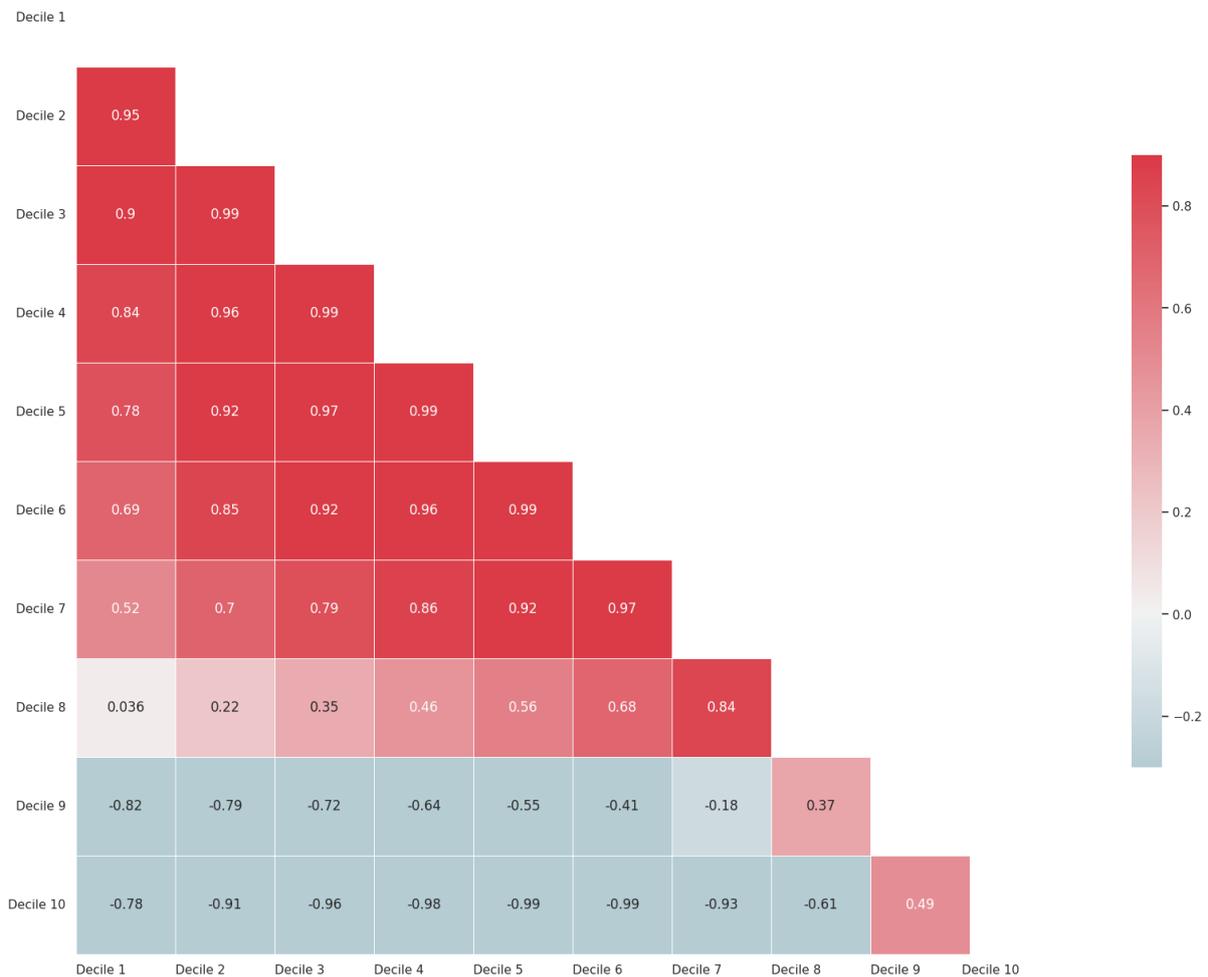

**Figure 6.3 Correlation between decile group income shares**
Source: own elaboration based on GCIP data.

We treat the last part of our analysis as a robustness test and apply additional panel regressions and AME for the shares of individual income deciles (see table 6.2 and online appendix), with an average model fit of 0.91. Overall, the dependence between variables was similar to the general correlation between them. Democratization did not positively affect the incomes of the bottom 40% of society. However, deciles from the fifth to the ninth benefited from democratization in the second period, though this effect was balanced by lower shares of income among the top 10%, where the impact was strongest. The most equalizing effect was observed with liberalization, which had a redistributive character favoring the bottom 80%. In contrast to the general models, the highest values of the rule of law did not favor equality, reducing the shares of income among the 90% of society without affecting the wealthiest decile. The previously mentioned educational rent most favorably affects the lowest deciles, improving their situation relative to the rest of the income distribution. Globalization brought gains mainly to the top 10%, in line with expectations. This part of the analysis also explained the counterintuitive impact of unemployment. For 90% of society, a negative coefficient direction was observed, and although not high, they were offset by benefits for the top 10%.

Additionally, the AME results showed that an increase in liberalization and greater opportunity for political expression is beneficial for the bottom 80% of incomes, reducing the incomes of the top 20%. Over the 2001-2016 period, liberalization benefits for the bottom 7 deciles ranged from 0.4 percentage point to 0.9 percentage point, with the highest effects accruing to the bottom half of the income distribution. These results were offset by a 0.5 percentage point decline in the income shares of the top 2 deciles. Meanwhile, the effects of higher regulation of participation mainly took on values equal to 0.1 percentage point lowering the shares of the 9th decile by 0.2 percentage point and the 10th decile by 0.7 percentage point Moreover, higher rule of law brings benefits to the seventh to tenth deciles equal to 0.1 percentage point, reducing the shares by the same amount for the bottom 30%. Moreover, the overall increase in democratization, especially in the later period, was most favorable for the deciles from the fifth to the ninth with 0.2-0.7 percentage point effects, primarily reducing the income shares of the highest decile at 1.3 percentage points and the three lowest shares at about 0.2-0.7. For the above analysis, it should be emphasized that when interpreting the results for individual deciles, it is important to take into account the scale of values shown in Table 4.1. For example, an effect of 0.5 percentage point is relatively high for the first decile, whose average share value over the entire period is 2.5%, and less significant for the tenth decile, with average share value of 29.9%

The results of panel regression and AME allowed us to establish that if a given aspect of democratization benefits 80% or 90% of people in the lower part of the income distribution, it is balanced by a reduction in the share of total income for the top one or two deciles. Nevertheless, benefits below the 9th decile do not always mean their simultaneous occurrence, much less equal in terms of their impact, for all lower parts of the distribution. However, this does not allow us to conclude that autocratic rule was more beneficial for these deciles. Especially considering that for the first, the lowest decile, where the rule of law does not improve the income situation, there is a benefit from the development of liberal democracy, which is linked to counteracting the tyranny of power, and benefit from a longer period of education, previously indicated as an indirect channel of democracy development; also the lowest category of the variable related to regulation of participation has the most adverse impact on the share of income of the poorest 10%. Ultimately, we agree with Wiseman (2017), finding that the development of democratization has brought benefits in terms of a more egalitarian distribution of income for at least 80% of those at the bottom of the distribution, with disproportionately lower returns, especially for the top decile.

**Table 6.2 Regressions for income shares by deciles**

| | P0-P10 | | P10-P20 | | P20-P30 | | P30-P40 | | P40-P50 | |
|---|---|---|---|---|---|---|---|---|---|---|
| | 1991-2000 | 2001-2016 | 1991-2000 | 2001-2016 | 1991-2000 | 2001-2016 | 1991-2000 | 2001-2016 | 1991-2000 | 2001-2016 |
| democratization scale | -0.004*** | -0.003* | -0.003*** | -0.002 | -0.003*** | -0.001 | -0.002* | 0.00002 | -0.001 | 0.001 |
| rule of law | -0.0003 | 0.003* | -0.002 | 0.002 | -0.002 | 0.001 | -0.003* | -0.0002 | -0.004* | -0.001 |
| liberal democracy | 0.015*** | 0.017*** | 0.021*** | 0.021*** | 0.023*** | 0.022*** | 0.024*** | 0.021*** | 0.022*** | 0.019*** |
| equal protection | -0.001 | -0.001 | -0.001 | -0.001 | -0.001 | -0.001 | -0.001 | -0.001 | -0.001 | -0.001 |
| regulation of participation [2] | -0.008* | 0.001 | -0.006 | 0.004 | -0.004 | 0.004* | -0.003 | 0.004 | -0.002 | 0.004 |
| regulation of participation [3] | -0.005 | 0.003 | -0.004 | 0.005* | -0.003 | 0.005* | -0.002 | 0.004 | -0.0003 | 0.004 |
| regulation of participation [4] | -0.01** | 0.002 | -0.008* | 0.006* | -0.006 | 0.006* | -0.004 | 0.006** | -0.002 | 0.006** |
| regulation of participation [5] | -0.002 | 0.000002 | -0.001 | 0.004 | -0.0002 | 0.005 | 0.0004 | 0.005 | 0.001 | 0.004 |
| GDP growth | 0.0001 | -0.000002 | -0.0000001 | -0.00001 | -0.00003 | -0.00002 | -0.0001 | -0.00001 | -0.0001 | 0.000001 |
| years of life | 0.004* | 0.001 | 0.005** | 0.003 | 0.007*** | 0.005 | 0.009*** | 0.007 | 0.011*** | 0.01** |
| years of education | 0.009* | 0.001 | 0.008* | 0.008* | 0.007 | 0.01* | 0.005 | 0.011* | 0.003 | 0.011* |
| export | -0.002 | 0.009 | -0.003 | 0.008 | -0.003 | 0.007 | -0.002 | 0.007 | -0.002 | 0.007 |
| import | -0.004 | -0.009 | -0.002 | -0.008* | -0.001 | -0.009* | -0.001 | -0.009* | -0.001 | -0.008* |
| globalization | -0.006 | 0.002 | -0.007* | -0.005 | -0.008* | -0.009 | -0.008* | -0.012 | -0.008* | -0.015* |
| unemployment | -0.0002* | 0.00001 | -0.0003** | -0.00006 | -0.0003** | -0.0001 | -0.0003** | -0.0002* | -0.0003*** | -0.0002** |
| Country RE | ✓ | ✓ | ✓ | ✓ | ✓ | ✓ | ✓ | ✓ | ✓ | ✓ |
| Period RE | ✓ | ✓ | ✓ | ✓ | ✓ | ✓ | ✓ | ✓ | ✓ | ✓ |
| Cov. Estimator | Robust | Robust | Robust | Robust | Robust | Robust | Robust | Robust | Robust | Robust |

| $R^2$ | 0.75 | 0.65 | 0.88 | 0.78 | 0.93 | 0.84 | 0.95 | 0.88 | 0.97 | 0.91 |
|---|---|---|---|---|---|---|---|---|---|---|
| | **P50-P60** | | **P60-P70** | | **P70-P80** | | **P80-P90** | | **P90-P100** | |
| | **1991-2000** | **2001-2016** | **1991-2000** | **2001-2016** | **1991-2000** | **2001-2016** | **1991-2000** | **2001-2016** | **1991-2000** | **2001-2016** |
| democratization scale | -0.001 | 0.002 | 0.0003 | 0.002* | 0.002 | 0.003*** | 0.003*** | 0.003** | 0.007 | -0.007* |
| rule of law | -0.004* | -0.003* | -0.004* | -0.004** | -0.003 | -0.006*** | -0.003 | -0.008*** | 0.006 | -0.006 |
| liberal democracy | 0.019*** | 0.016*** | 0.013*** | 0.01*** | 0.003 | 0.003 | -0.016*** | -0.013*** | -0.134*** | -0.121*** |
| equal protection | -0.001 | -0.0004 | -0.001 | -0.002 | -0.001 | 0.0003 | -0.0003 | 0.0004 | 0.009 | 0.005 |
| regulation of partisanship [2] | 0.0003 | 0.003 | 0.003 | 0.002 | 0.006** | 0.001 | 0.01*** | -0.0001 | 0.015 | 0.002 |
| regulation of partisanship [3] | 0.001 | 0.003 | 0.003 | 0.002 | 0.005* | 0.001 | 0.009*** | -0.001 | 0.003 | -0.002 |
| regulation of partisanship [4] | 0.001 | 0.005* | 0.004 | 0.004 | 0.008*** | 0.002 | 0.015*** | -0.002 | 0.024 | -0.011 |
| regulation of partisanship [5] | 0.002 | 0.003 | 0.004 | 0.002 | 0.006* | 0.0003 | 0.008*** | -0.002 | -0.004 | -0.003 |
| GDP growth | -0.000001 | 0.00001 | -0.0001 | 0.00002 | -0.0001 | 0.00004 | -0.0001 | 0.0001 | -0.0001 | 0.001 |
| years of life | 0.014*** | 0.012*** | 0.016*** | 0.015*** | 0.02*** | 0.018*** | 0.022*** | 0.019*** | 0.011 | 0.018 |
| years of education | -0.0001 | 0.01* | -0.004 | 0.006 | -0.008* | 0.001 | -0.015*** | -0.009* | -0.012 | -0.061* |
| export | -0.001 | 0.006 | -0.0001 | 0.006 | 0.001 | 0.005* | 0.003 | 0.002 | 0.015 | -0.064* |
| import | -0.001 | -0.008* | -0.002 | -0.006* | -0.002 | -0.004* | -0.001 | 0.001 | 0.008 | 0.048* |
| globalization | -0.008* | -0.016* | -0.008* | -0.016* | -0.006* | -0.012* | 0.001 | 0.004 | 0.063** | 0.104* |
| unemployment | -0.0002** | -0.0002*** | -0.0002* | -0.0002*** | -0.0001 | -0.0002*** | 0.0001* | -0.0001* | 0.002* | 0.002** |
| Country RE | ✓ | ✓ | ✓ | ✓ | ✓ | ✓ | ✓ | ✓ | ✓ | ✓ |
| Period RE | ✓ | ✓ | ✓ | ✓ | ✓ | ✓ | ✓ | ✓ | ✓ | ✓ |
| Cov. Estimator | Robust | Robust | Robust | Robust | Robust | Robust | Robust | Robust | Robust | Robust |
| $R^2$ | 0.98 | 0.95 | 0.98 | 0.97 | 0.98 | 0.99 | 0.98 | 0.99 | 0.93 | 0.85 |

Signif. codes: 0.001 '***' 0.01 '**' 0.05 '*'
Source: own elaboration based on data presented in section 4.1.

**Conclusions**

In the article, we examined the relation between democratization and income inequality in post-socialist countries between 1991 and 2016. The aim was to find answers to two hypotheses. The first hypothesis focused on determining whether there was a relationship between democratization and income inequality in the analyzed countries during the initial period of economic transition, characterized by intense changes in many areas. The second hypothesis aimed at determining whom democratization benefited if there was a co-occurence of this process with income inequality in at least some time period. Our main analysis involved with panel regressions and average marginal effects. The analysis was conducted for the Palma ratio, Gini coefficient, and shares of total income for individual income deciles as part of robustness testing. In addition, the study was divided into two sub-periods due to the supposition of time lags in the effects of democratization on the distribution of income. The division was set at the year 2000, which, based on the literature and our analysis, considered the end of the intense changes associated with the transition. Based on the findings presented, several conclusions can be drawn.

Firstly, due to statistically significant results for the main variables related to democracy and control variables related to the economy and quality of life, we find a co-occurence of both political and economic institutions channels with the level of income inequality in the studied group of countries. Our findings suggest that improvements in the overall index of democratization, stronger rule of law, equal protection of all social groups, and liberal democracy contribute to higher levels of income equality at an aggregated level. However, the variables associated with democratization had a more frequent and stronger impact on the values of the Palma ratio and Gini coefficient in the second period studied. Moreover, the effects of single variables shown by the AME analysis for the years 2001-2016 indicate the possibility of contributing to a reduction in the value of the Palma ratio by up to at least 4.1% in the case of an increase in the scale of democratization by one unit. In the case of the Gini coefficient, pro-equality effects ranged from 0.5 to 1.1 percentage points, which, given the occurrence of simultaneous changes in several channels of inequality impact, may significantly translate into a lower level of income inequality in the population. In contrast, the same analysis showed that for the 1991-2000 period, the democratic variables were mainly characterized by positive effects on measures of inequality, meaning they increased the overall level of income inequality in society.

We conclude that the development of democracy had a particular correlation with a reduction in income inequalities after 2000. We find support for this conclusion throughout the entire analysis. In contrast, although in the first sub-period several measures in the panel regressions indicated the presence of some relation or co-occurrence, we suggest that this effect was apparent. During the initial transition period, there was an increase in both income inequality and democratization indices, which is also indicated by the results of the AME, mainly identifying the effects of the increase in inequality by democratization at that time. Any relation between these two processes is difficult to study before the 21st century due to the period of transition and the many simultaneous changes in the economies under study. Moreover, as we pointed out earlier, the rise in inequality in the 1990s was a natural process resulting directly from a change in the system that had previously kept it artificially low. It was thus a process isolated from the changes taking place at the level of democratization and not a genuinely interconnected process. Only when inequality began to stabilize, i.e., reached its true, natural levels, are we able to determine whether this relationship actually occurred, which happened only in the second sub-period studied, where the results indicate that the relationship is pro-equality. All of these results allow us to confirm both hypotheses.

Additionally, we wanted to examine who actually gained during the period when we found a relationship between democratization and the level of income inequality. In pursuance of

Wiseman (2017) as well as Roine and Waldenström (2015), we demonstrated that democratization was beneficial in terms of having a higher share in total income for at least 80% of society from the lower part of the income distribution, adversely affecting especially the incomes of the top 10% or, in some cases, 20%. However, a more equal income distribution does not necessarily mean equal benefits for all bottom deciles. Moreover, we do not find arguments in favor of the thesis suggested in the literature by Beitz (1991) and Förster & Tóth (2015) about the possibility of higher benefits for the poorest in authoritarian systems, based on the example of post-socialist countries. Moreover, we find confirmation for the redistributive effect of extended years of education, which co-occurred with the development of new systems and favoritism towards the highest decile in terms of benefiting from globalization.

In conclusion, our study allowed us to confirm both hypotheses formulated. It shows that between 1991 and 2000, the actual relationship between democratization and income inequality did not exist, or at most was illusory. On the other hand, between 2001 and 2016, the relationship between those two processes was present and had a pro-equality character. During that period, the development of the democratic system benefited at least 80% of the lower part of the income distribution, at the expense especially of the top decile's share of total income. These results confirmed that democratization positively affects the shares of lower income deciles in post-socialist countries.

In the article we only studied the relationship in the form of the impact of democratization on income distribution and did not split up the group of countries studied. This is because the direction of the relationship we focused on is mainly discussed in this area of research and the problem applies to all post-socialist countries, by which we tried to find a solution that would explain why the whole group is problematic in terms of empirical research. This approach is a certain limitation of our analysis and given the results obtained in subsequent studies. Literature indicates that the channels through which democratization can affect inequality are the spread of voting rights and higher citizen participation in influencing government, which can affect decision-making, ensuring a more equitable distribution of income. Similarly, policies aimed at reducing the disparity of human capital, mainly associated with democratic governments should also have a positive impact, as well as increased redistribution, aimed at the welfare of the majority. Studies also point to the positive effects of higher tax rates, increased civil liberties and generally higher inclusiveness of systems. In order to confirm the detailed significance of the indicated factors and to deepen the conclusions we have identified, it is important to conduct research that does not aim to examine the co-occurrence of two processes or the existence of their relationship, but to analyze causality. Importantly, it is relevant not only to study the impact of democratization on income inequality but also to examine the inverse relationship - the impact of changes in income inequality on the development of democratic systems.